\documentclass[runningheads]{llncs}
\usepackage{graphicx}
\usepackage{caption}
\usepackage{subcaption}

\begin{document}
\title{Rate Adaptation Aware Positioning for Flying Gateways using Reinforcement Learning}
\titlerunning{RA Aware Positioning for Flying Gateways using RL}
%
\author{Gabriella Pantaleão\orcidID{0009-0007-0255-572X} \and
Rúben Queirós\orcidID{0000-0001-7804-0246} \and
Hélder Fontes\orcidID{0000-0002-7672-8335} \and
Rui Campos\orcidID{0000-0001-9419-6670}}
\authorrunning{Gabriella Pantaleão et al.}
%
\institute{INESC TEC and Faculdade de Engenharia, Universidade do Porto, Portugal
\email{\{gabriella.pantaleao, ruben.m.queiros, helder.m.fontes, rui.l.campos\}@}inesctec.pt}
\maketitle              
\begin{abstract}
With the growing connectivity demands, Unmanned Aerial Vehicles (UAVs) have emerged as a prominent component in the deployment of Next Generation On-demand Wireless Networks. However, current UAV positioning solutions typically neglect the impact of Rate Adaptation (RA) algorithms or simplify its effect by considering ideal and non-implementable RA algorithms. This work proposes the Rate Adaptation aware RL-based Flying Gateway Positioning (RARL) algorithm, a positioning method for Flying Gateways that applies Deep Q-Learning, accounting for the dynamic data rate imposed by the underlying RA algorithm. The RARL algorithm aims to maximize the throughput of the flying wireless links serving one or more Flying Access Points, which in turn serve ground terminals. The performance evaluation of the RARL algorithm demonstrates that it is capable of taking into account the effect of the underlying RA algorithm and achieve the maximum throughput in all analysed static and mobile scenarios.

\end{abstract}
\section{Introduction}
Unmanned Aerial Vehicles (UAVs), commonly known as drones, have emerged as a promising technology for ensuring cost-effective on-demand wireless connectivity \cite{bs_optimization,wireless_uav}. The ability of UAVs to fly and navigate autonomously allow them to provide access to communications services where no infrastructure coverage exists, which can be particularly useful in remote or difficult-to-reach areas. UAVs are especially relevant in the context of sudden fluctuations in traffic demands that impair the effective allocation of radio resources, a scenario frequently seen in crowded events, for instance. Thus, UAVs are suitable platforms for delivering and enhancing connectivity in heterogeneous scenarios, carrying communications hardware to deploy Wi-Fi or cellular coverage. UAVs are highly adaptable and can be quickly deployed and optimally positioned, in contrast to conventional ground-based solutions \cite{uav_placement_survey}. 

\begin{figure}[ht]
     \centering
    \includegraphics[width=0.5\textwidth]{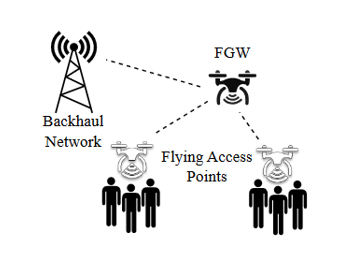}
    \caption[]{Relay network topology example, adapted from \cite{queiros2023trajectory}.}
\label{fig:topology}
\end{figure}

In flying networks, relays are often employed to improve network coverage and capacity, as represented in Figure~\ref{fig:topology}. The distance between UAVs directly influences the communications range between the flying nodes, as well as the Quality of Service (QoS) associated to the inter-UAV wireless links. Therefore,
the use of intermediate UAVs, herein named Flying Gateways (FGWs), enables the extension of the communication range between the Backhaul network and the Flying Access Points (FAPs), while preserving the quality of the established connection \cite{positioning_andre_coelho}. By optimizing the positioning of the FGWs, there is an effective balance of the load in the network while improving coverage, resulting in a more efficient and reliable flying network. Nevertheless, the optimization of the position of UAVs in flying networks according to a specific link metrics is still a challenge, considering their dynamic nature. 

Machine Learning approaches can be employed to enhance various aspects of the performance and operation of flying networks, including delay, throughput, transmission power, and cache resource utilization \cite{positioning_dlr_eduardo,drl_trajectory_bs,rl_multiple_uav,rl_uav_flowlevel,drl_uav_mobile_edge}. Reinforcement Learning (RL) techniques are particularly useful, as they are intrinsically related to control theory. RL represents a relevant approach to handle the continuous changing environment of flying networks, taking into account long-term, sequential and cumulative rewards, as it maps states into the effectiveness of the actions \cite{rl_an_overview}. With RL, decisions are based on the current states of the environment, meaning that the agent can be trained using real-time network measurements instead of relying on approximations \cite{positioning_dlr_eduardo}. This allows the agent to have a more accurate response to the current environment conditions.

The main contribution of this paper is the Rate Adaptation aware RL-based Flying Gateway Positioning (RARL) algorithm, which addresses the problem of finding the position for the FGW that maximizes the throughput obtained in the FAPs, considering the influence of the underlying RA algorithm. The effectiveness of the RL approach was validated through simulations using Network Simulator 3 (ns-3), which served as the platform for training, validating and testing the model. 

The rest of the paper is organized as follows. Section II presents the related work on UAV positioning in flying networks. The design of the RARL algorithm is detailed in Section III. Section IV introduces the ns-3 simulation setup, as well as the simulation results on the performance of the RARL algorithm. Finally, Section V presents the conclusions and points out the future work. 

\section{Related Work}
With the advancements made in the context of flying networks, many studies have been developed to improve the positioning of the flying nodes. The work carried in \cite{positioning_andre_coelho} includes the implementation of centralized on-demand Gateway UAV positioning algorithm, relying on the awareness of the incoming traffic behaviour. Based on a mathematical approach, the goal is to minimize the capacity of the bidirectional wireless link between the FGW and the FAPs, while ensuring the bitrates required by each FAP. The results indicate that the position of the FGW is an essential aspect of the Backhaul network configuration, since constant FGW position updates can improve network performance. However, the authors consider an ideal underlying RA algorithm. 

In \cite{drl_trajectory_bs}, a solution for flying Base Stations, serving a region with numerous users, was proposed. This deployment relies on RL techniques for the UAV to be able to learn how to optimize its trajectory, contouring obstacles and reaching the intended service area. In this study, the UAV is assumed to return to its landing location when reaching the limit period of the flight. With a Q-Learning approach, the system was implemented based on making direct movement decisions, allowing the optimization of the sum rate of the transmission over the duration of the flight. With this solution, the algorithm needs no prior knowledge of the environment and has the ability to learn the structure of the network, resulting in improved performance. However, a limiting factor of this study is that it considers a static environment, with no changes in the course of the flight.

Overall, the main UAV positioning approaches include brute-force searching, mathematical optimizations \cite{positioning_andre_coelho,3d_placement_uav_bs,joint_traj_power}, heuristics \cite{heuristics}, and Machine Learning algorithms \cite{positioning_dlr_eduardo,drl_trajectory_bs,rl_multiple_uav,rl_uav_flowlevel,drl_uav_mobile_edge}. Despite their importance, optimizing UAV positions in a network to maximize one or more link metrics remains a challenging task, given the multiple factors to consider. From the literature review, DRL emerges as a promising approach for UAV positioning, supporting the choice of the Q-Learning approach in the implementation of the RARL algorithm. 

It is worth noting that few solutions have considered the impact of the Backhaul network on the QoS experienced by ground terminals \cite{positioning_andre_coelho}. In addition, they are focussed on static and invariant scenarios, assuming mostly symmetric links. Moreover, none of the studies analysed included a realistic RA algorithm, which limits the applicability of these methods. Our work aims to address the aforementioned shortfalls, by considering a continuous analysis of the network state and accounting for the implementation of realistic and dynamic adjustments of the data rate in the network.

\section{RARL Algorithm Design}
\begin{figure}[h!]
\centering
    \begin{subfigure}{0.45\textwidth}
     \centering
    \includegraphics[width=\textwidth]{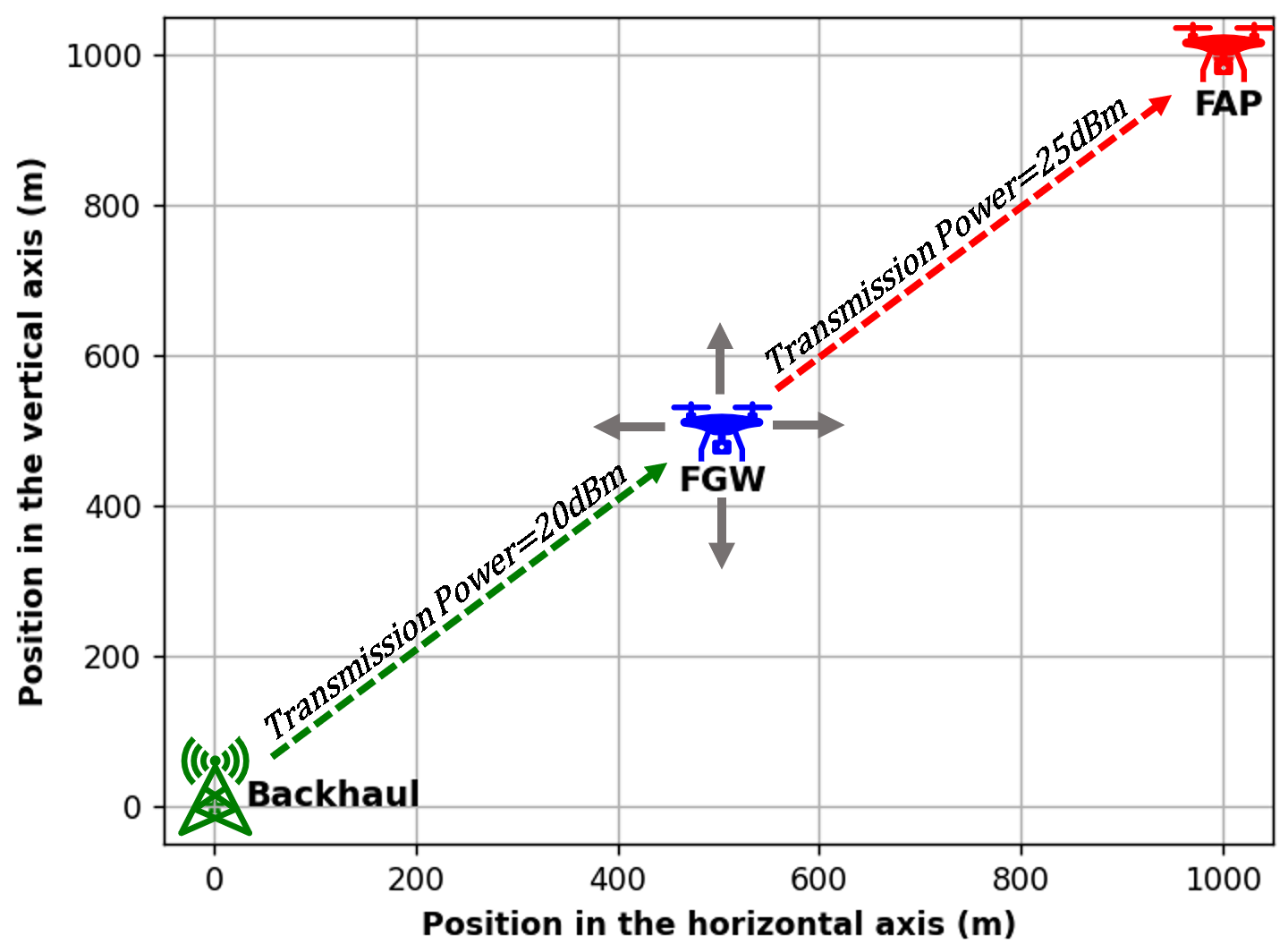}
    \caption{\textbf{Asymmetric links} scenario.}
    \label{fig:sc1}
     \end{subfigure}
\hfill
     \begin{subfigure}{0.45\textwidth}
     \centering
    \includegraphics[width=\textwidth]{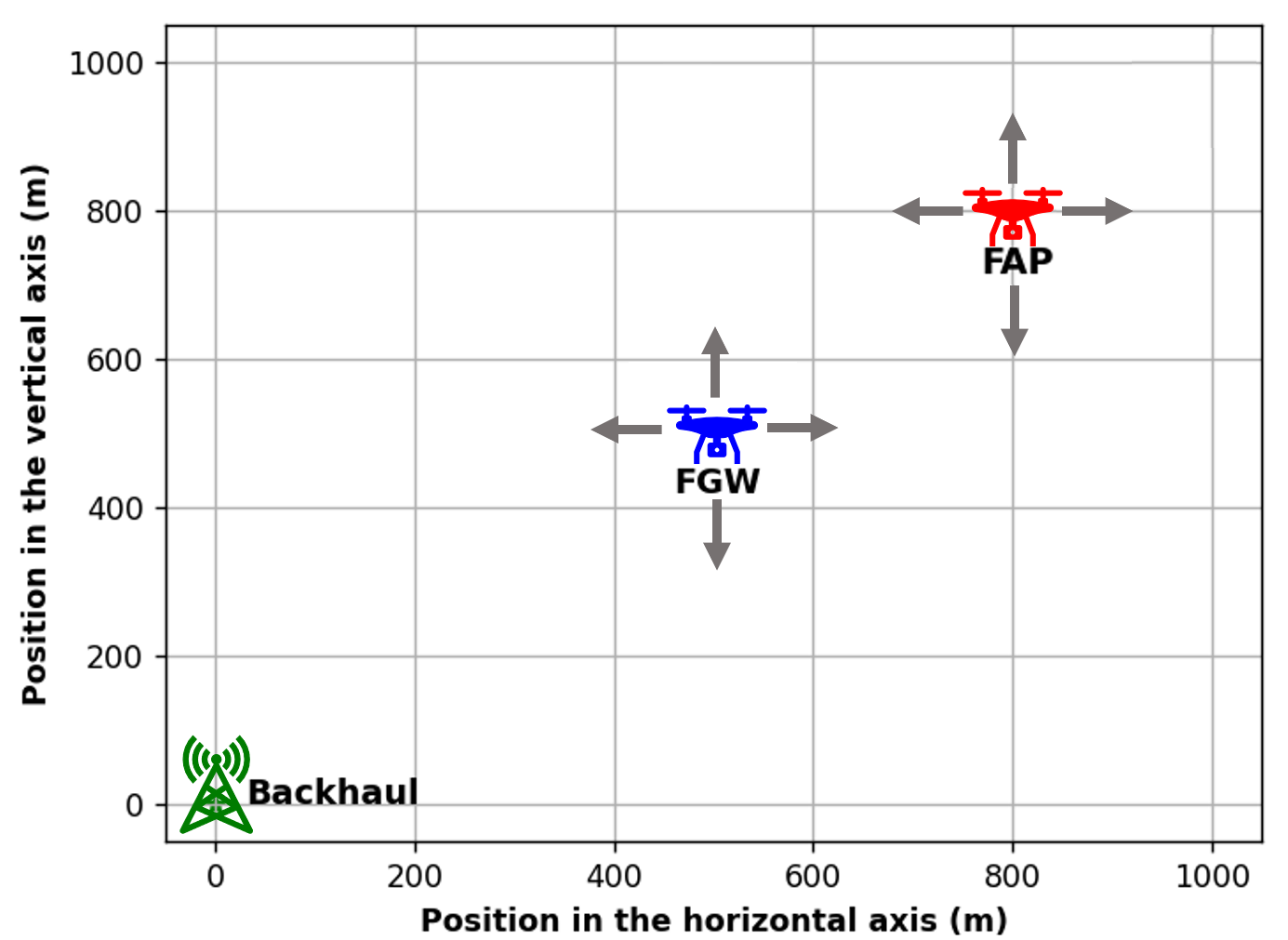}
    \caption{\textbf{Moving FAP} scenario.}
    \label{fig:sc2}
     \end{subfigure}
     \hfill
     \begin{subfigure}{0.45\textwidth}
     \centering
    \includegraphics[width=\textwidth]{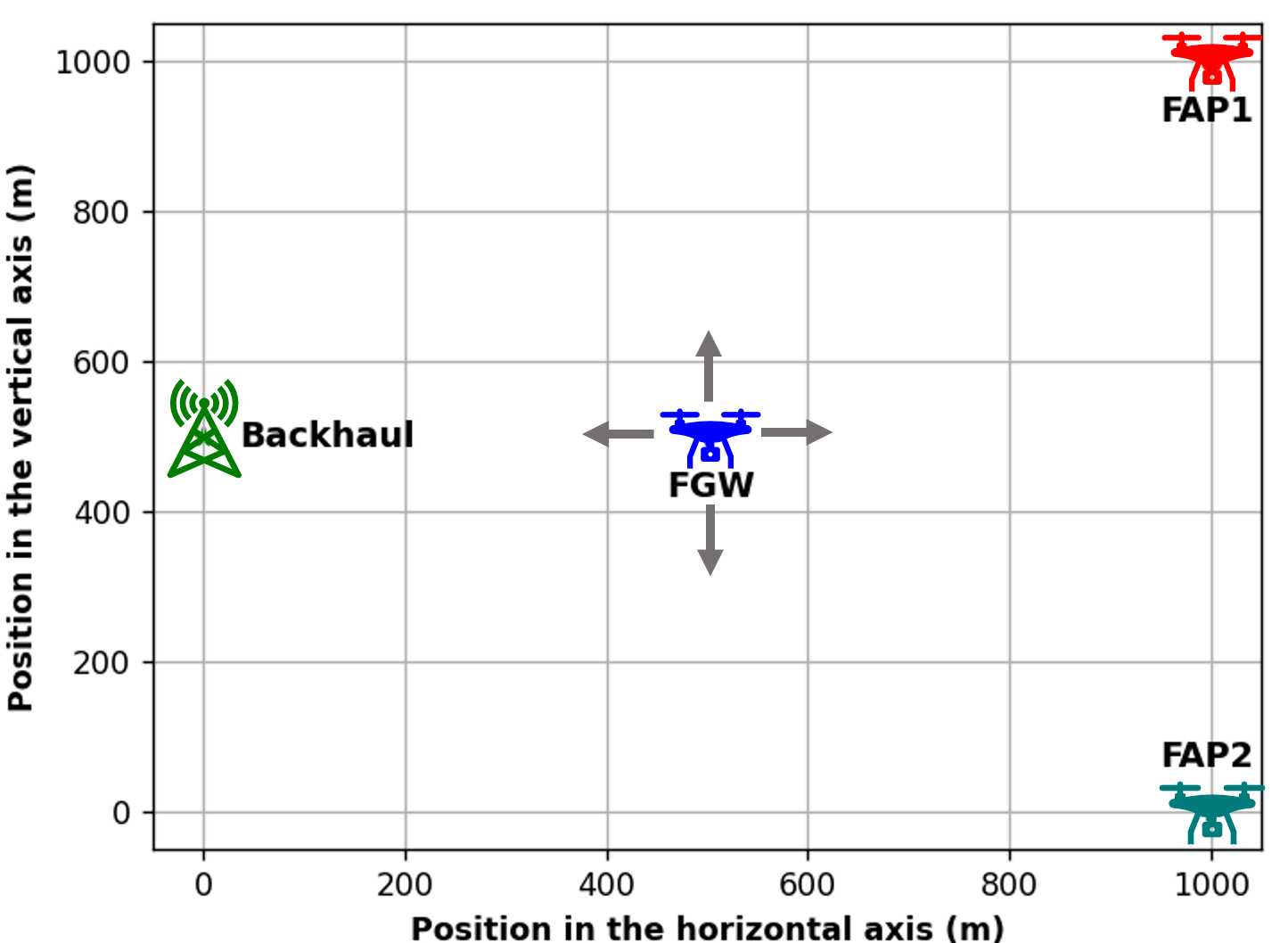}
    \caption{\textbf{Two FAPs} scenario.}
    \label{fig:sc3}
     \end{subfigure}
\hfill
\caption{Three scenarios analysed.}
\label{fig:scenarios}
\end{figure}

The RARL algorithm relies on the implementation of Deep Q-Learning, a Deep RL method that has been proven to be highly applicable in continuous control tasks, making it well-suited for addressing the positioning of UAVs within flying networks. Three different scenarios were defined to design the RARL algorithm, as presented below:
\begin{enumerate}
    \item \textbf{Asymmetric links scenario: }with three static nodes corresponding to the Backhaul, the FGW and the FAP, considering that the link between the FGW and the Backhaul node and the link between the FAP and the FGW are asymmetric, meaning that they have different values for the transmission power. This scenario is illustrated in Figure~\ref{fig:sc1}.
    \item \textbf{Moving FAP scenario: }with the same network topology as the \textbf{asymmetric links} scenario, where the link between the FAP and the FGW and the link between the FGW and the Backhaul node are symmetric, but the FAP moves, as presented in Figure~\ref{fig:sc2}.
    \item \textbf{Two FAPs scenario: }with four static nodes, corresponding to the Backhaul, the FGW and two FAPs, where the link between the FGW and the Backhaul node handles the traffic of the two symmetric links between the FGW and the FAPs, as shown in Figure~\ref{fig:sc3}.
\end{enumerate}

Without loss of generality, in the two last scenarios we have considered symmetric wireless links in order to isolate the problems. The learning process was adapted for each of the three scenarios. For all the cases, the agent was instantiated in the FGW node, which is responsible for performing the different actions selected. Nevertheless, the algorithm modelling as a Markov Decision Process was different for each scenario. The action space, the observation space and the reward functions defined for each scenario are described below.

\subsection{Action Space}
The actions of the FGW are based on discrete sequential movements, based on incrementing the position in the 2D venue in the selected direction by 25 meters, a distance defined as a compromise between the training time and the impact of the FGW's movement on the link metrics, selected every 1 second, defined as the decision interval. Thus, the action space comprises five movement options, including the possibility of remaining in the same position: \textit{Left, Right, Up, Down, } and \textit{Same}.

\subsection{Observation Space}
The observations must characterize the system so that the agent can recognize the best positions for maximizing and balancing the throughput values in the FGW and FAP nodes. For the \textbf{asymmetric links} and \textbf{moving FAP} scenarios, the observations are defined as:

\begin{itemize}
    \item the coordinates of the FGW;
    \item the distances between the FGW and the Backhaul node and between the FGW and the FAP;
    \item the throughput values measured in each link, calculated as the number of bytes received throughout the course of a decision interval, i.e. every second. 
\end{itemize}

For the \textbf{two FAPs} scenario, the approach to sense the environment is different, given the existence of more than one FAP. In this case, the observations do not include the throughput values.

\subsection{Reward Function}
The underlying Minstrel-HT RA algorithm has a significant impact on the obtained throughput values, heavily influenced by the distances travelled by the FGW. This means that even if it travels to the same final position, the initial conditions imposed by the link metrics in the starting point impact the final throughput value obtained at the destination point. This means that the closer the UAV originally is to the final position, the fewer fluctuations are seen in the observed throughput. This is due to the fact that, when using Minstrel-HT, the throughput variation has a slow response in cases where the
link quality improves, as shown in \cite{dara,rateAdapAlgos,minHT,minstrel_eval}. Hence, modelling the system becomes challenging due to the need to identify positions that optimize and balance the throughput values, which in turn requires addressing the peaks and variations introduced by the Minstrel-HT algorithm. 

To evaluate the variation of the link metrics - in this case the Signal-to-Noise Ratio (SNR) and the throughput - throughout the simulations, a preliminary study was carried out. For this purpose, the Backhaul node was positioned in \textit{(0, 0)} and the FAP in \textit{(1000, 1000)}, with the configuration showed in Figure~\ref{fig:sc1}. By moving the FGW from the coordinates \textit{(25, 25)} to \textit{(975, 975)} and incrementing the position of the FGW, horizontally and vertically, by 25 meters every 1 second, the aim was to analyse the actual impact of the distance in the link metrics. 
To evaluate the link metrics variation throughout the simulations, a preliminary study was carried out. For this, the Backhaul node was positioned in \textit{(0, 0)} and the FAP in \textit{(1000, 1000)}, with the configuration showed in Figure~\ref{fig:sc1}. By moving the FGW from the coordinates \textit{(25, 25)} to \textit{(975, 975)} and incrementing the position of the FGW, horizontally and vertically, by 25 meters every 1 second, the aim was to analyse the actual impact of the distance in the link metrics. 

\begin{figure}[ht]
\centering
    \begin{subfigure}{0.45\textwidth}
     \centering
    \includegraphics[width=\textwidth]{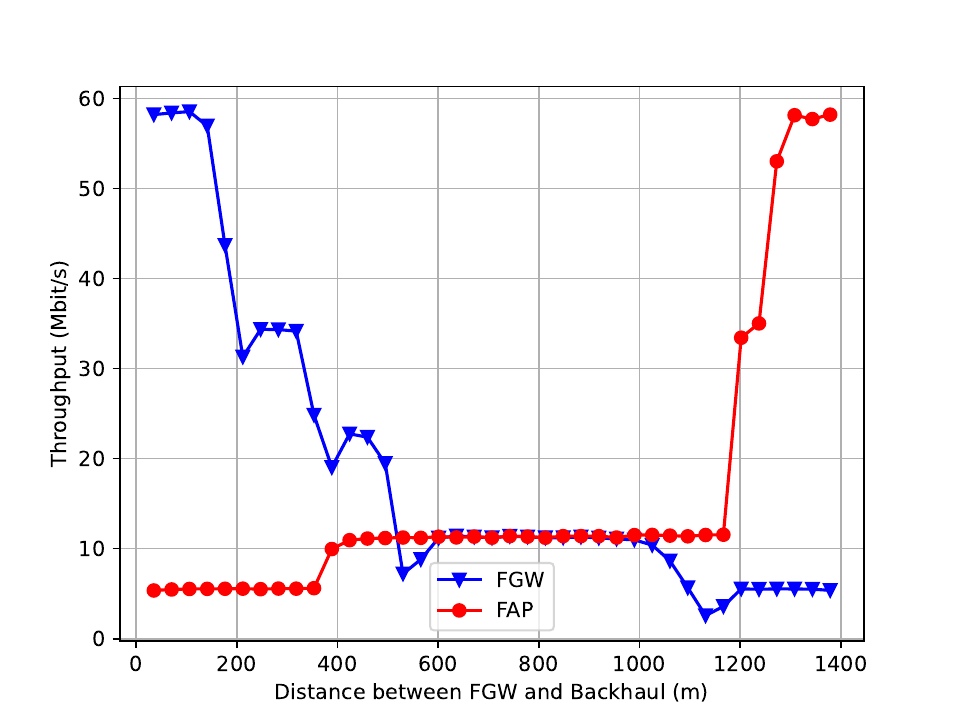}
    \caption{Throughput variation in nodes FAP and FGW.}
    \label{fig:tgp1}
     \end{subfigure}
\hfill
     \begin{subfigure}{0.45\textwidth}
     \centering
    \includegraphics[width=\textwidth]{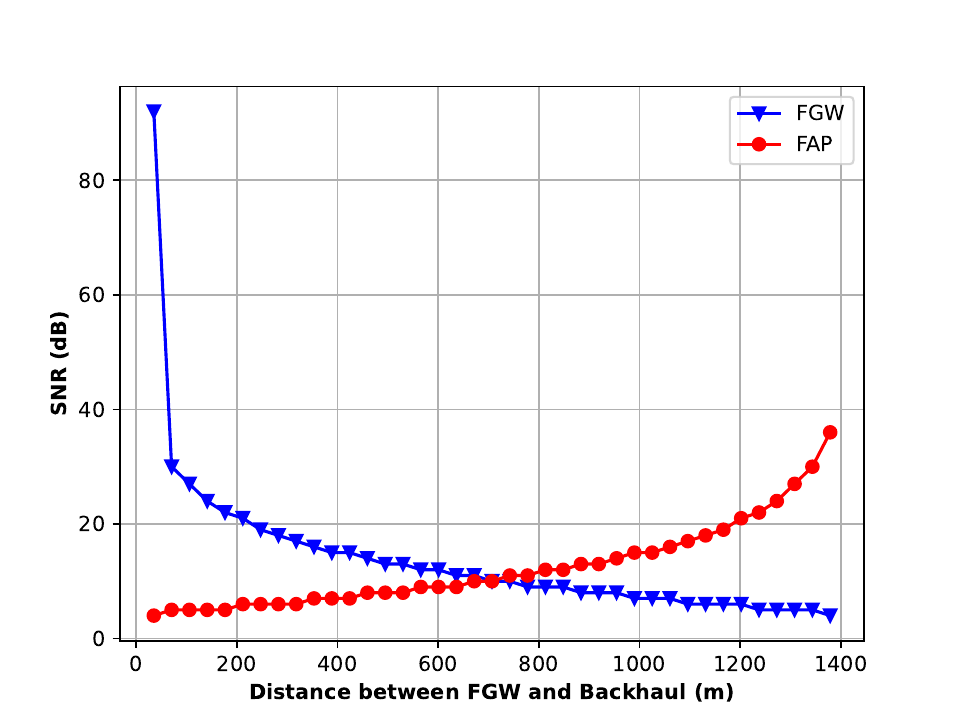}
    \caption{SNR variation in nodes FAP and FGW.}
    \label{fig:snr1}
     \end{subfigure}
\hfill
\caption{Link metrics comparison for the \textbf{asymmetric links} and \textbf{moving FAP} scenarios.}
\label{fig:reward1}
\end{figure}

For the \textbf{asymmetric links} and \textbf{moving FAP} scenarios, since the SNR, shown in Figure~\ref{fig:snr1}, has greater sensitivity to changes in position when compared to the throughput, presented in Figure~\ref{fig:tgp1}, the SNR was considered as a more suitable measure of the environment’s state. This means that, the objective can be modelled according to the SNR, with the reward functions translating into maximizing the SNR values and minimizing the difference between the SNR obtained in both the link connecting the Backhaul node and the FGW and in the link between the FGW and the FAP. 

The goal is to maximize the throughput values and to minimize possible throughput imbalances between the two nodes acting as receivers in each wireless link. For this purpose, the objective is then to maximize the SNR value in the FGW, $SNR_{FGW}$, and the SNR value in FAP, $SNR_{FAP}$, and penalize the difference between them, by multiplying it by a weight of 2. This constant was chosen empirically, as it allows the penalization to be sufficient to impact the reward value when the links are imbalanced, without impacting negatively the learning process. The reward function is defined in Equation~\ref{eq:eq2}:

\begin{equation} \label{eq:eq2}
 Reward = SNR_{FGW} + SNR_{FAP} - 2|SNR_{FGW} - SNR_{FAP}| 
\end{equation} 

\begin{figure}[h!]
\centering
    \begin{subfigure}{0.45\textwidth}
     \centering
    \includegraphics[width=\textwidth]{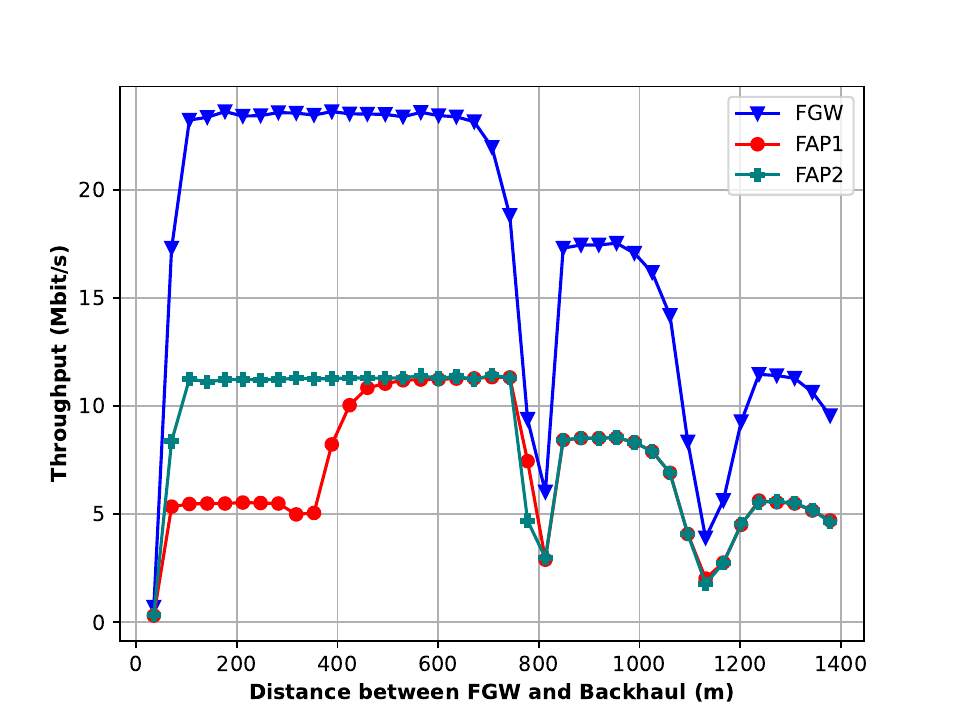}
    \caption{Throughput variation in FAPs and FGW.}
    \label{fig:tgp2}
     \end{subfigure}
\hfill
     \begin{subfigure}{0.45\textwidth}
     \centering
    \includegraphics[width=\textwidth]{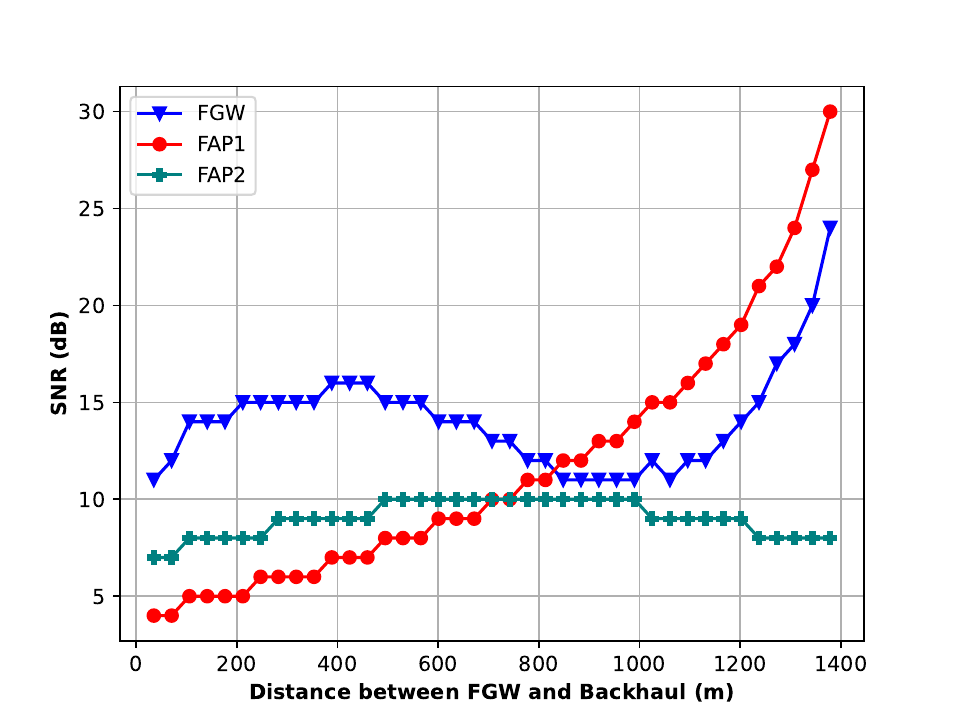}
    \caption{SNR variation in FAPs and FGW.}
    \label{fig:snr2}
     \end{subfigure}
\hfill
\caption{Link metrics comparison for the \textbf{two FAPs} scenario.}
\label{fig:reward2}
\end{figure}

In the case of the \textbf{two FAPs} scenario, the objective remains the same: to maximize the throughput values while minimizing throughput imbalances between the two nodes. However, due to differences in topology and the need for routing traffic through the FGW node, the variations of the throughput and SNR values are different in this scenario, given the increase of the Packet Loss Ratio as the distance between the FGW and the Backhaul node increases, which implies a reduction of the link capacity. This is demonstrated by the results obtained from the same preliminary study conducted for this scenario, where the Backhaul node was positioned in \textit{(0, 500)}, the FAP1 in \textit{(1000, 1000)} and  the FAP2 in \textit{(1000, 0)}, as represented in Figure~\ref{fig:sc3}. The results of the throughput variation in Figure~\ref{fig:tgp2} and the SNR variation in Figure~\ref{fig:snr2} show that the throughput and SNR are not proportional, due to the interdependency of the links. For instance, if we move the FGW closer to the Backhaul node, the links between the FGW and the FAPs get stretched, thus leading to a reduction in the throughput obtained in the link between the FGW and the Backhaul node. 

Hence, in this scenario, the reward was designed to rely directly on the throughput values obtained in the FAPs. As in the first reward function, the reward was designed to penalize the possible throughput imbalances obtained in the FAPs and favour higher throughput values. Similarly, a factor of 2 was considered. The reward function for the \textbf{two FAPs} scenario is defined in Equation~\ref{eq:eq3}, where \textit{T} refers to the throughput.

\begin{equation} \label{eq:eq3}
 Reward = T_{FAP_1} + T_{FAP_2} - 2|T_{FAP_1} - T_{FAP_2}|
\end{equation} 

\section{Simulation Configuration and Results} \label{sec:obtained}
The simulations performed to validate the proposed RARL algorithm were carried out using ns-3. This section presents the configuration of the simulation environment and the simulation results obtained for each scenario. 

\subsection{Simulation Configuration}
The ns-3 simulator played a crucial role in simulating real-world behaviour in various scenarios. It served as the foundation for the DRL agent to gather relevant information from the environment, specifically the link metrics of interest. It is worth noting that the different connections were designed as independent links and downlink traffic was considered. Table~\ref{tab:ns3-gen} summarizes the most relevant parameters applied in the simulation.

\begin{table}[h!]
\centering
\caption{Simulation parameters considered for the three scenarios.}
\begin{tabular}[ht]{ p{9cm} p{2.5cm} }
 \hline
 \multicolumn{2}{c}{\textbf{ns-3.37 Simulator Parameters}} \\
 \hline
 Wi-Fi Standard& IEEE 802.11n \\
 Channel Bandwidth& 20 MHz \\
 Antenna Gain& 0 dBi \\ 
 Propagation Loss Model& Friis \\
 Rate Adaptation Algorithm& Minstrel-HT \\
 Application Traffic& UDP \\
 UDP Data Rate& 70 Mbit/s \\
 Packet Size& 1400 bytes \\
  \hline
\end{tabular}
\label{tab:ns3-gen}
\end{table}

\subsection{Simulation Results}
The initial positions of the FGW were defined as extreme locations, tailored to each specific scenario. The results are presented below according to the scenario.

\subsubsection{Asymmetric Links Scenario.}
Regarding the \textbf{asymmetric links} scenario, the aim was to assess the impact of the asymmetric links in the optimal positioning of the FGW. Thus, the Backhaul and the FAP nodes were stationary, located at the coordinates (0, 0) and (1000, 1000), respectively. Furthermore, the initial position for the FGW was at the coordinates (500, 500), to show that the model was able to recognize the asymmetry between the links and find the optimal position in the new scenario, following the trajectory shown in Figure~\ref{fig:500_500_min}, as depicted below:
\begin{itemize}
    \item Initial position of the FGW: (500, 500);
    \item Final position of the FGW: (175, 525);
    \item Final distance between FGW and FAP: 950 meters;
    \item Final distance between FGW and Backhaul: 550  meters.
\end{itemize}

Regarding the \textbf{asymmetric links} scenario, it is possible to see that the FGW recognizes the asymmetry of the links, as the final position is significantly closer to the Backhaul node, given that its transmission power was lower than the transmission power of the FGW. After a few iterations, the throughput in the FAP and in the FGW reached the value of around 17 Mbit/s, as seen in Figure~\ref{fig:500_500_tgps1}. Nevertheless, the instabilities with the Minstrel-HT RA algorithm were evident, with fluctuations occurring during the Modulation and Coding Scheme data rate changes.

Figure~\ref{fig:500_500_min} also presents the baseline solution, with the optimal trajectory to the optimal final position, determined as the point that ensures the SNR values in both links are the same, following Equation~\ref{eq:eq4}, where \emph{P} refers to the transmission power, \emph{G} to the antenna gain, \emph{D} to the distance between nodes and \emph{f} to the frequency of operation. The throughput variation for the baseline solution, shown in Figure~\ref{fig:500_500_tgps2}, follows a pattern similar to the one achieved with the RARL algorithm. Still, RARL converges faster (cf. Figure 5b), validating the RA aware implementation of the RARL algorithm, as it was able to quickly meet the objective of throughput converge and maximization in the different links.


\begin{equation} \label{eq:eq4}
P_{Rx}= P_{Tx} + G_{Tx} + G_{Rx} + 20\log\left(\frac{c}{{4\pi Df}}\right)
\end{equation} 

\begin{figure}[h!]
\centering
    \begin{subfigure}{0.44\textwidth}
     \centering
    \includegraphics[width=\textwidth]{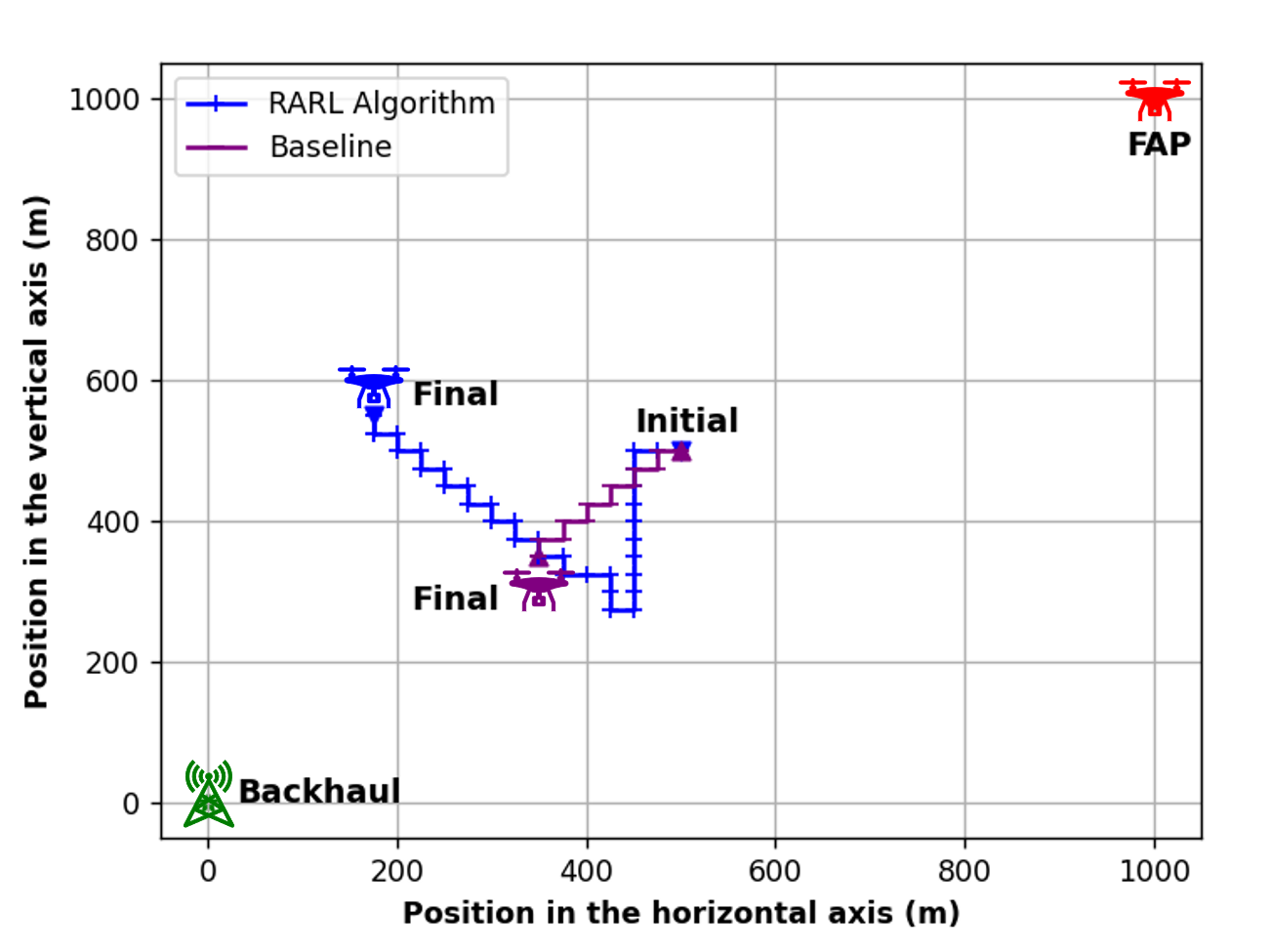}
    \caption{Trajectory of the FGW to final position.}
    \label{fig:500_500_min}
     \end{subfigure}
\hfill
     \begin{subfigure}{0.45\textwidth}
     \centering
    \includegraphics[width=\textwidth]{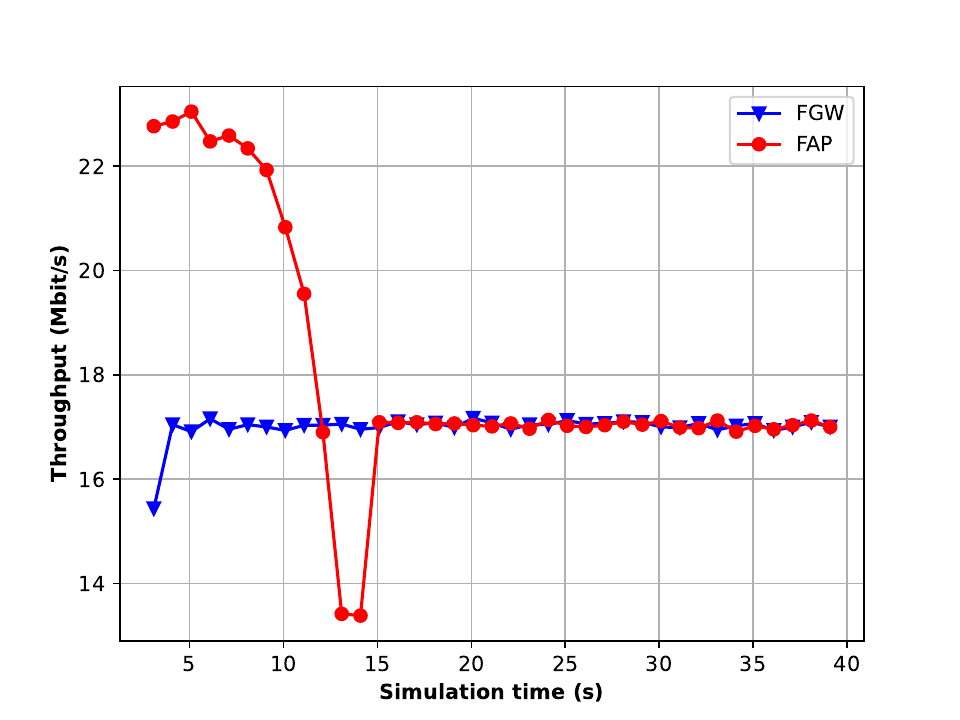}
    \caption{Throughput values evolution for the RARL algorithm.}
    \label{fig:500_500_tgps1}
     \end{subfigure}
\hfill
\hfill
     \begin{subfigure}{0.45\textwidth}
     \centering
    \includegraphics[width=\textwidth]{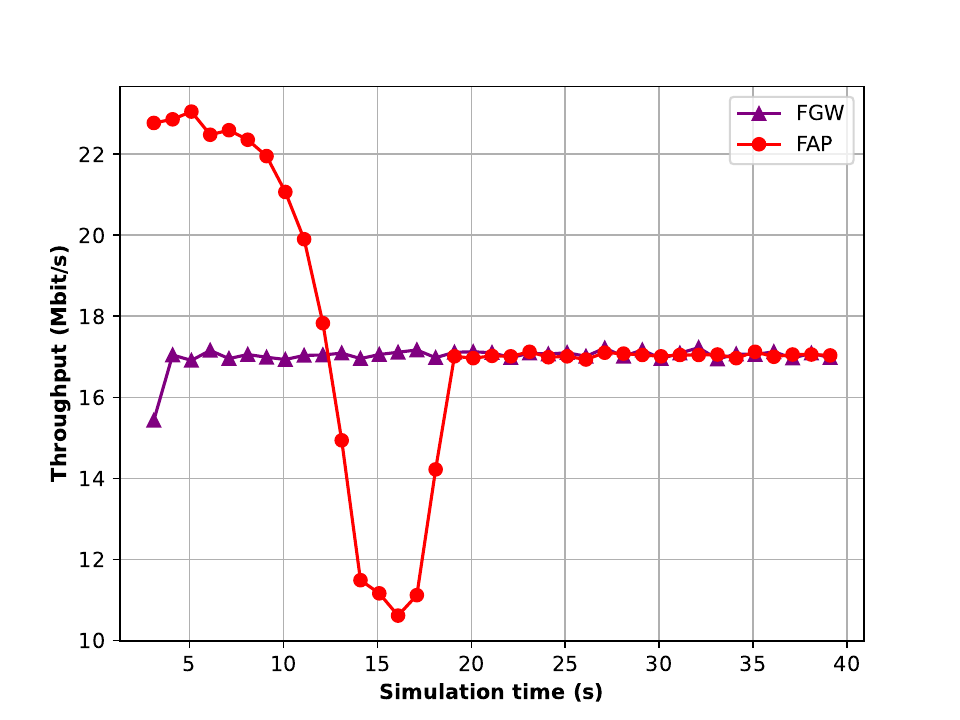}
    \caption{Throughput values evolution for the baseline model.}
    \label{fig:500_500_tgps2}
     \end{subfigure}
\hfill
\caption{Analysis of the RARL algorithm performance for the \textbf{asymmetric links} scenario.}
\label{fig:asymm_min}
\end{figure}

\subsubsection{Moving FAP Scenario.}
\begin{figure}[h!]
\centering
    \begin{subfigure}{0.45\textwidth}
     \centering
    \includegraphics[width=\textwidth]{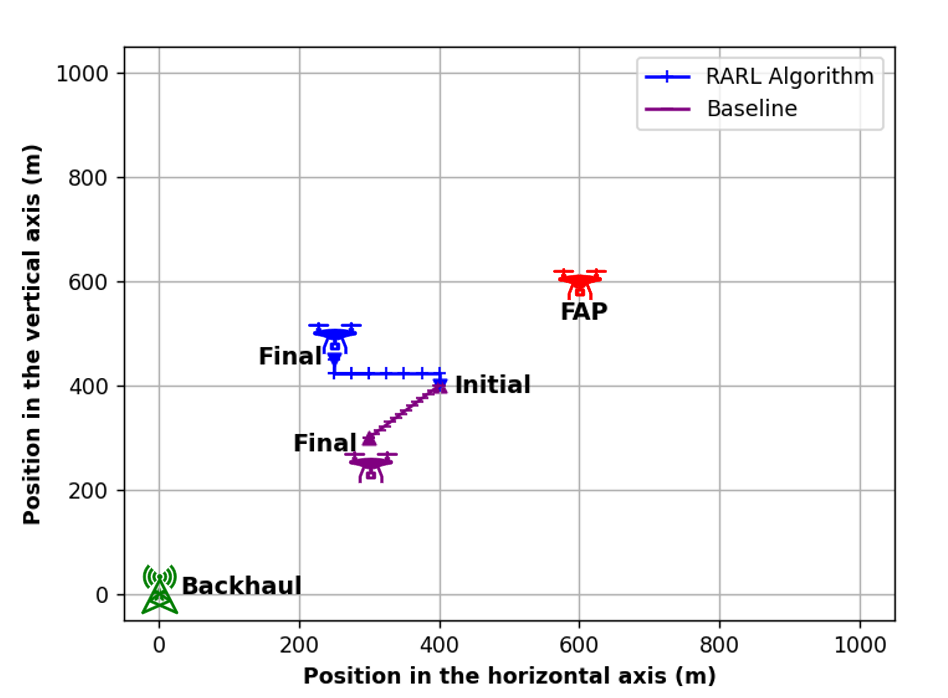}
    \caption{First part of the movement.}
    \label{fig:move1}
     \end{subfigure}
\hfill
     \begin{subfigure}{0.45\textwidth}
     \centering
    \includegraphics[width=\textwidth]{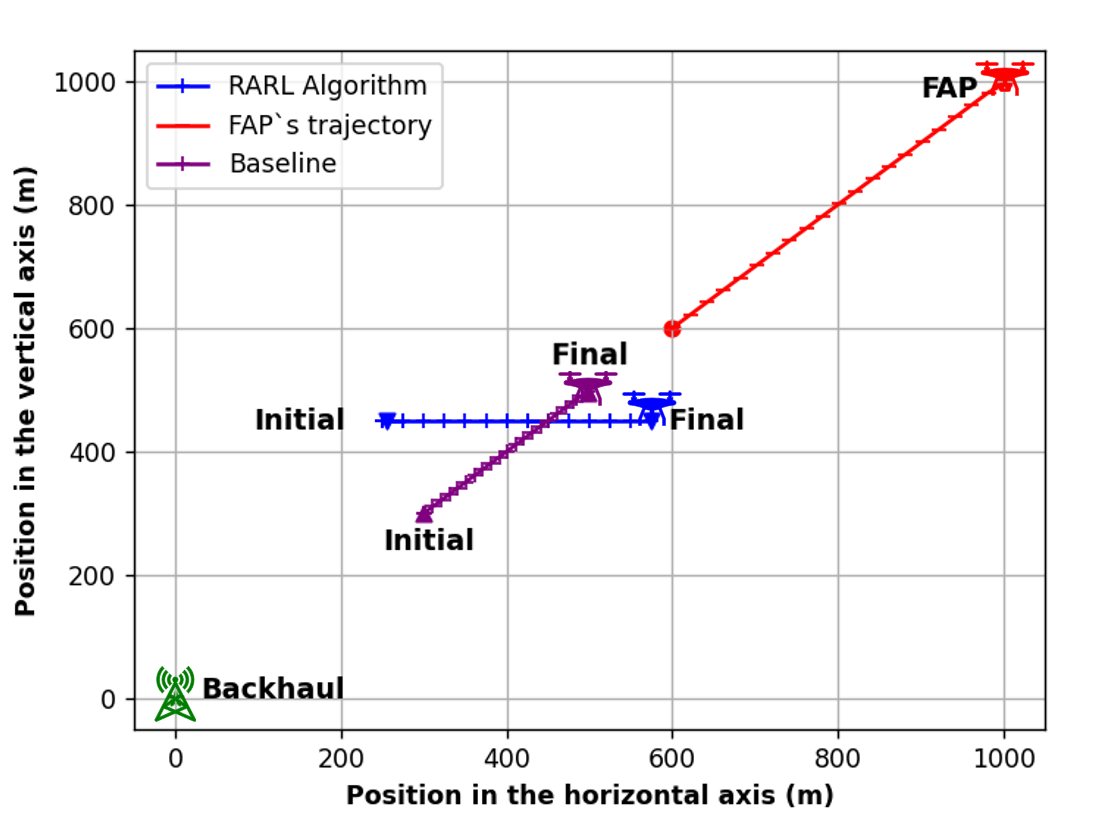}
    \caption{Second part of the movement.}
    \label{fig:move2}
     \end{subfigure}
\hfill
\begin{subfigure}{0.45\textwidth}
     \centering
    \includegraphics[width=\textwidth]{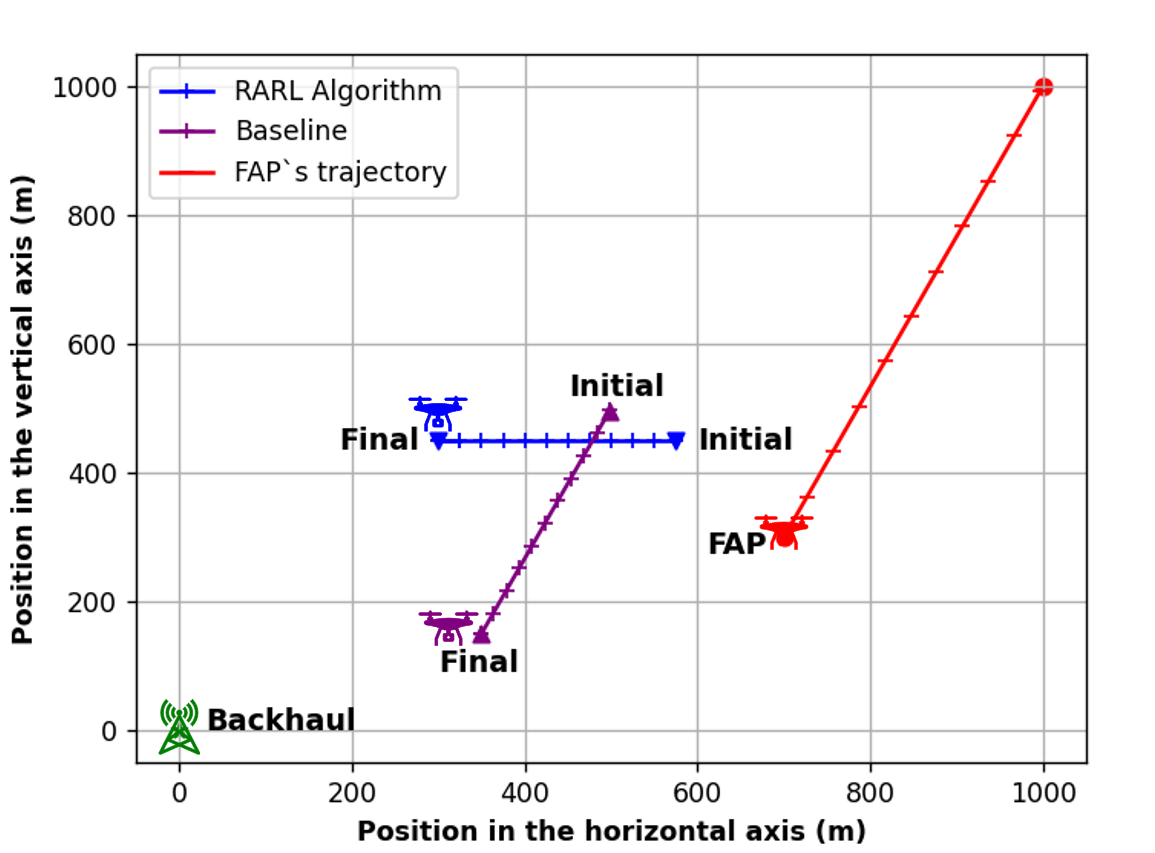}
    \caption{Third part of the movement.}
    \label{fig:move3}
     \end{subfigure}
\caption{Trajectory of the FGW to final position for the \textbf{moving FAP} scenario.}
\label{fig:moves}
\end{figure}

The FAP movement in the \textbf{moving FAP} scenario relied on the use of the Waypoint Mobility Model available in ns-3. For the evaluation presented herein, the movement was defined to evidence the behaviour regarding the FGW positioning when the FAP moves closer and further away. The following movement was defined:
\begin{enumerate}
    \item The FAP is stationary during the first Waypoint, at (600, 600), for the initial 20 seconds, as observed in Figure~\ref{fig:move1}. The final results are:
    \begin{itemize}
        \item Initial position of the FGW: (400, 400);
        \item Final position of the FGW: oscillation between the coordinates (250, 450) and (275, 450);
        \item Final distance between FGW and FAP: 380 meters;
        \item Final distance between FGW and Backhaul node: 500 meters.
    \end{itemize} 
    \item When it comes to the second Waypoint, represented in Figure~\ref{fig:move2}, the movement of the FAP from (600, 600) to (1000, 1000). The FGW behaviour is described below:
    \begin{itemize}
        \item Initial position of the FGW: (275, 450);
        \item Final position of the FGW: the FGW adapts progressively to the FAPs position, finally oscillating between (550, 450) and (575, 450). The final position of this trajectory validates the RARL algorithm, given the proximity to the geometric centre between the FAP and the Backhaul node;
        \item Final distance between FGW and FAP: 700 meters;
        \item Final distance between FGW and Backhaul node: 730 meters.
    \end{itemize} 
    \item Finally, in the third part of the movement, displayed in Figure~\ref{fig:move3}, the FAP moves to the coordinates (700, 300). The FGW behaviour is described below:
    \begin{itemize}
        \item Initial position of the FGW: (575, 450);
        \item Final position of the FGW: the final position of the FGW resulted in the oscillation between the coordinates (300, 450) and (325, 450). 
        \item Final distance between FGW and FAP: 400 meters;
        \item Final distance between FGW and Backhaul node: 550 meters.
    \end{itemize} 
\end{enumerate}

\begin{figure}[h!]
\centering
     \begin{subfigure}{0.45\textwidth}
     \centering
    \includegraphics[width=\textwidth]{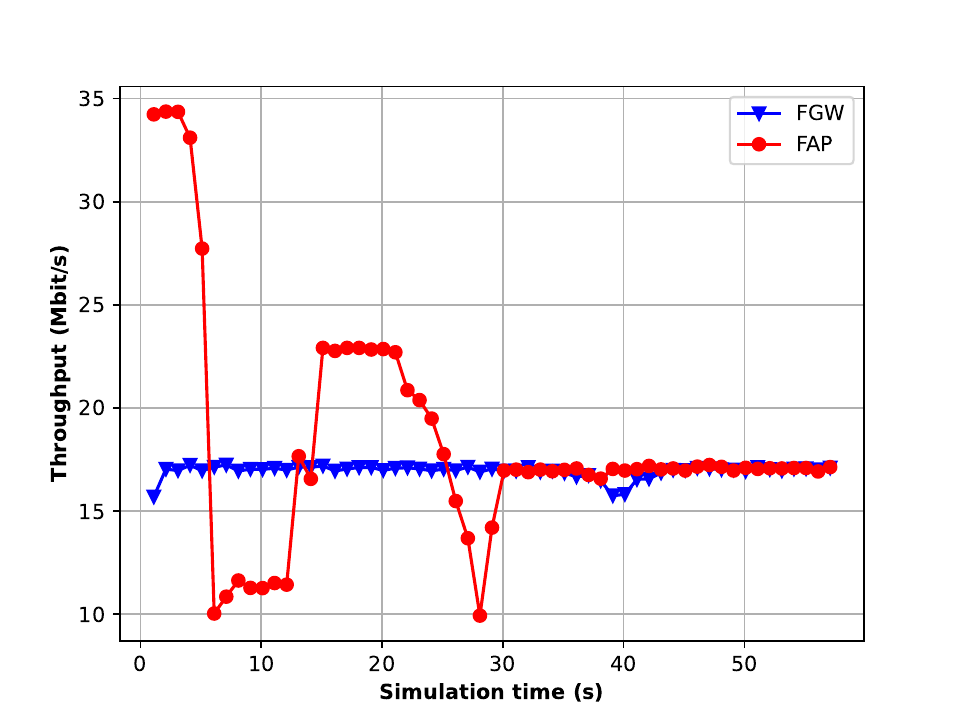}
    \caption{Throughput values evolution for the RARL algorithm.}
    \label{fig:tgps_move}
     \end{subfigure}
\hfill
\hfill
     \begin{subfigure}{0.45\textwidth}
     \centering
    \includegraphics[width=\textwidth]{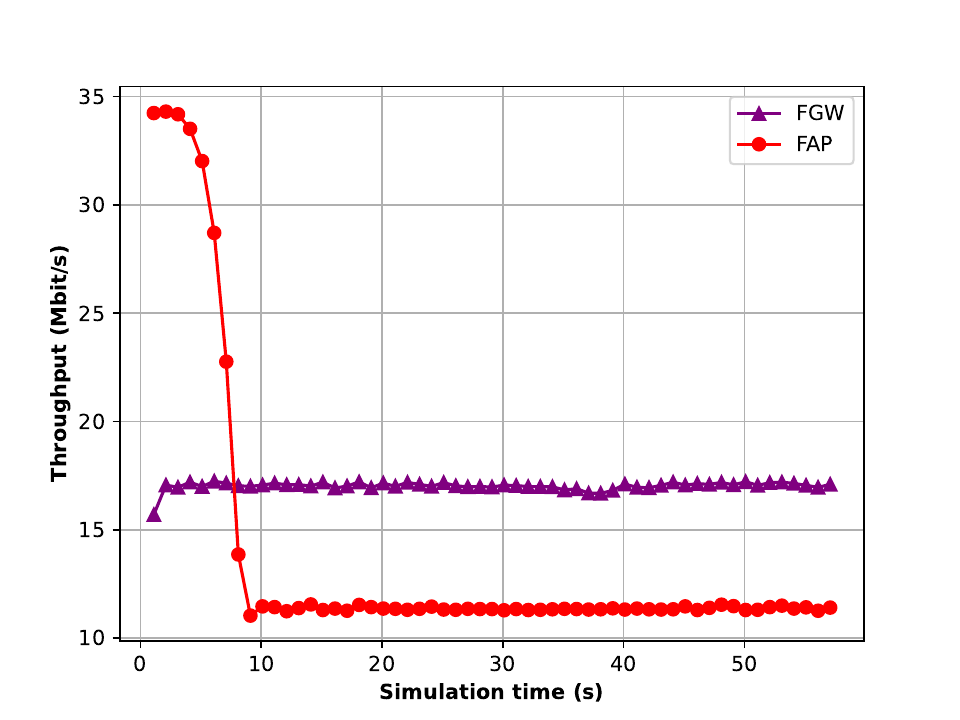}
    \caption{Throughput values evolution for the baseline model.}
    \label{fig:tgps_move2}
     \end{subfigure}
\hfill
\caption{Throughput values evolution for the \textbf{moving FAP} scenario.}
\label{fig:moves2}
\end{figure}

When it comes to the evolution of the throughput values throughout the simulation in the \textbf{moving FAP} scenario, it is possible to observe in Figure~\ref{fig:tgps_move} that the throughput in the FGW node remained overall constant, having a uniform behaviour. On the other hand, the throughput measured in the FAP suffered multiple variations. Given that the FAP movement begins at 20 seconds, the fluctuations occurred just in the initial moments. As soon as the FGW is able to adjust the position to the movement, the throughputs converge, reaching around 17 Mbit/s. Due to the convergence occurring during the movement of the FAP, it can be concluded that the FGW was able to successfully follow the trajectory of the FAP.

A baseline model was also tested to evaluate the RARL algorithm performance. In the baseline model, the FGW defines a trajectory that follows the movement of the FAP node, maintaining the same distance between the FAP and the Backhaul node. The throughputs in the receiving nodes are presented in Figure~\ref{fig:tgps_move2} and lead to the conclusion that the RARL was able to learn how to converge and maximize the throughput values even when the environment is constantly changing. As an RA aware algorithm, the RARL algorithm was able to detect the throughput imbalances and ensure the convergence of the throughputs throughout the FAP, outperform the baseline solution.

\subsubsection{Two FAPs Scenario.}
\begin{figure}[h!]
\centering
    \begin{subfigure}{0.45\textwidth}
     \centering
    \includegraphics[width=\textwidth]{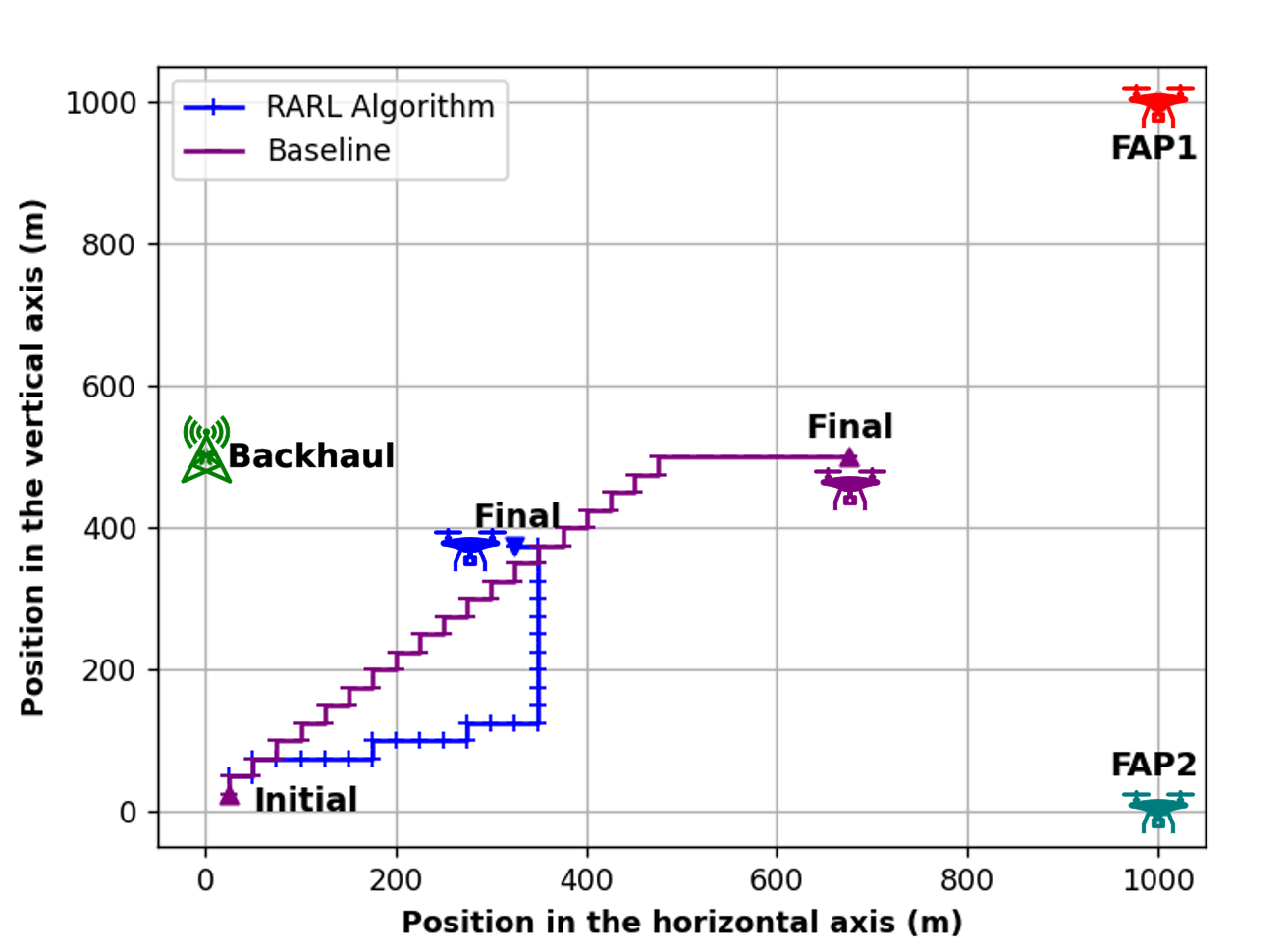}
    \caption{Trajectory of the FGW to the final position.}
    \label{fig:25_25_positions}
     \end{subfigure}
\hfill
     \begin{subfigure}{0.46\textwidth}
     \centering
    \includegraphics[width=\textwidth]{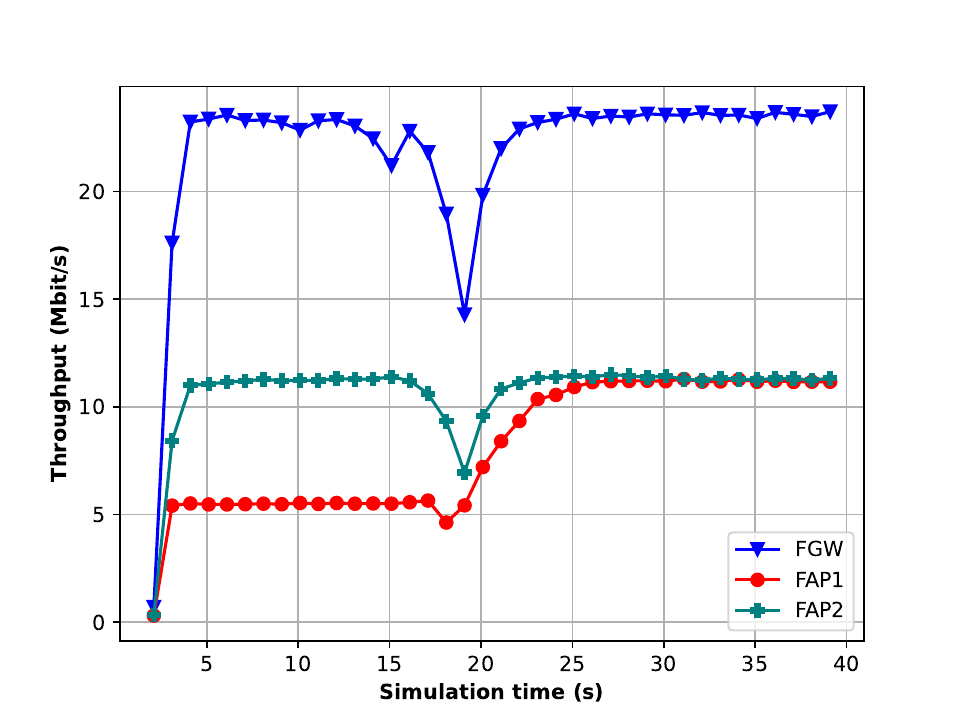}
    \caption{Throughput values evolution for the RARL algorithm.}
    \label{fig:tgps_multi}
     \end{subfigure}
\hfill
\hfill
     \begin{subfigure}{0.46\textwidth}
     \centering
    \includegraphics[width=\textwidth]{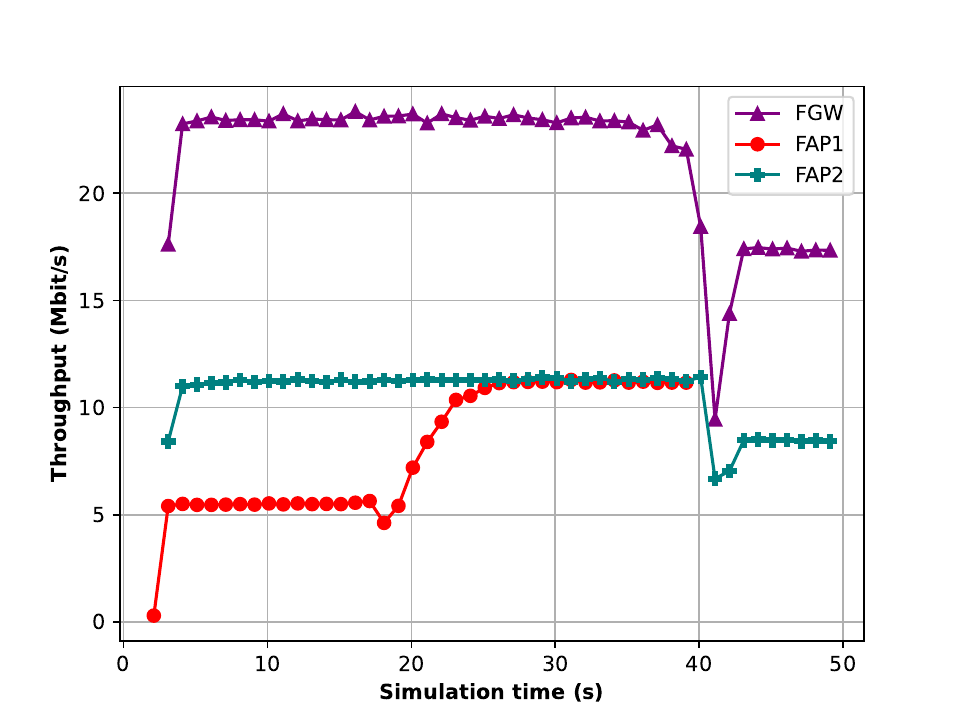}
    \caption{Throughput values evolution for the baseline model.}
    \label{fig:tgps_multi2}
     \end{subfigure}
\hfill
\caption{Analysis of the RARL algorithm performance for the \textbf{two FAPs} scenario.}
\label{fig:multi_min}
\end{figure}

When it comes to the \textbf{two FAPs} scenario, given the presence of an additional FAP node, the distribution of the nodes through the venue included the Backhaul node at (0, 500), FAP1 at (1000, 1000) and FAP2 at (1000, 0). The analysis of the implementation with the Minstrel-HT, shown in Figure~\ref{fig:25_25_positions}, was based on placing the FGW at a considerable distance to all the nodes in this configuration, having the following results:

\begin{itemize}
    \item Initial position of the FGW: (25, 25);
    \item Final position of the FGW: (325, 375);
    \item Final distance between FGW and FAP1: 920 meters;
    \item Final distance between FGW and FAP2: 770 meters;
    \item Final distance between FGW and Backhaul: 350 meters.
\end{itemize} 

Overall, in the \textbf{two FAPs} scenario, the first link between the Backhaul node and the FGW can act as a bottleneck, limiting the maximum throughput achievable in the FAPs. The throughput variation in the different nodes is shown in Figure~\ref{fig:tgps_multi}, showcasing the effect of the applied underlying RA algorithm with the instabilities associated with data rate transitions. It is possible to observe that the throughput in the FGW is constantly higher, showing values around 23 Mbit/s. However, around the moment when there is a transition to a higher data rate in FAP1, there is a significant fluctuation in all the values. Finally, the throughput values in the FAPs converge to approximately 11 Mbit/s, accomplishing the objective of reaching higher values. 

A baseline optimal trajectory leading to the geometric centre of the three nodes is presented in Figure~\ref{fig:25_25_positions}. This implementation was able to converge the throughput values throughout most of the simulation time. Nevertheless, the throughputs eventually decrease, which evidences that the final position was not optimal, showcasing that the RARL algorithm indeed maximized the throughput values in the nodes, while minimizing imbalances.

\section{Conclusions}
Given the great impact on the network performance, the optimal positioning of the FGW is a critical element in the flying network design. This paper proposes the RARL algorithm, a DRL-based FGW positioning approach that considers the effect of two aspects that have been overlooked in the state of the art: the influence of underlying RA algorithms and the impact of the Backhaul network configuration.

The evaluation of the performance of the RARL algorithm was carried out in ns-3. It was possible to observe that the fluctuations and instabilities, associated with the influence of Minstrel-HT in the link metrics, were overcome. This is supported by the trajectory of the FGW leading to a maximization of the defined reward functions, despite potential interference caused by the underlying RA algorithm in the throughput measured throughout the displacement of the FGW. 

The comparisons of the RARL algorithm with the baseline models demonstrate its capability of converging and maximizing the throughput values in the nodes in all the three scenarios studied, meaning that this work endorses the possibility of implementing an RA aware positioning algorithm for real-world deployments. As future work, we intend to enhance the trajectory performed by the FGW, as it was overall not fully optimized, in order to reduce resource consumption and the time to converge to the optimal position. 
%
%
%
\bibliographystyle{splncs04}
\bibliography{samplepaper}
\end{document}